\documentclass[preprint,amsmath,amssymb,superscriptaddress]{revtex4-1}
\usepackage{xcolor}
\usepackage{tabu}
\usepackage{titlesec}
\usepackage[utf8]{inputenc}
\usepackage{graphicx,epsfig,epstopdf}
\usepackage{bm}
\usepackage{epstopdf}
\usepackage{amssymb,amsmath,amsfonts,latexsym}
\usepackage{dcolumn}
\usepackage{bm}
\usepackage{hyperref}
\usepackage[utf8]{inputenc}
\newcommand{\nn}{\nonumber\\}
\usepackage{breqn}             
\makeatletter                  
\let\cat@comma@active\@empty   
\makeatother                   
\begin{document}
\title{Spectroscopy of heavy-heavy flavour mesons and annihilation widths of quarkonia}%

\author{Nakul R. Soni}
\email{nrsoni-apphy@msubaroda.ac.in}
\affiliation{Department of Physics, Faculty of Science,  \\ The Maharaja Sayajirao University of Baroda, Vadodara 390002, Gujarat,  India.}

\author{Rikita M. Parekh}
\affiliation{Department of Physics, Faculty of Science,  \\ The Maharaja Sayajirao University of Baroda, Vadodara 390002, Gujarat,  India.}

\author{Janaki J. Patel}
\affiliation{Department of Physics, Faculty of Science,  \\ The Maharaja Sayajirao University of Baroda, Vadodara 390002, Gujarat,  India.}
\affiliation{Applied Physics Department, Faculty of Technology and Engineering, \\ The Maharaja Sayajirao University of Baroda, Vadodara 390001, Gujarat,  India.}

\author{Akshay N. Gadaria}
\affiliation{Applied Physics Department, Faculty of Technology and Engineering, \\ The Maharaja Sayajirao University of Baroda, Vadodara 390001, Gujarat,  India.}

\author{Jignesh N. Pandya}
\email{jnpandya-phy@spuvvn.edu}
\affiliation{Department of Physics, Sardar Patel University,  Vallabh Vidyanagar 388120, Gujarat,  India.}

\date{\today}

\begin{abstract}
Within the framework of nonrelativistic quark-antiquark Cornell potential model formalism, we study the annihilation of heavy quarkonia. We determine their annihilation widths resulting into digluon, dilepton, $3\gamma$, $3g$ and $\gamma gg$ and compare our findings with the available theoretical results and experimental data.  We also provide the charge radii and absolute square of radial Schr\"odinger wave function at zero quark-antiquark separation for heavy quarkonia and $B_c$ mesons.
\end{abstract}

\maketitle
\section{Introduction}
\label{intro}
Heavy quarkonia and $B_c$ mesons are the bound state of a heavy quark and a heavy antiquark that can be theoretically treated using nonrelativistic formalism.
Decay properties of these states provide information regarding the internal structure as well as dynamics of the bound states.
With the advancements in experimental facilities world-wide, new results for the mass spectra as well as the decay properties continue to pour in for the heavy quarkonia sector.
Recently, LHCb Collaboration \cite{Aaij:2020jrx} provided the most precise measurement of ground state mass of $B_c$ meson. Also, the excited state mass was reported by CMS Collaboration in their experimental data \cite{Sirunyan:2019osb}. These results read,
\begin{eqnarray}
m(B_c^+) &=& 6274.47 \pm 0.27 \pm 0.17 ~\mathrm{MeV} \nonumber \\
m(B_c^+(2S)) &=& 6871.0 \pm 1.2 \pm 0.8 \pm 0.8 ~\mathrm{MeV}\nonumber
\end{eqnarray}
Very recently CMS collaboration presented the first ever simultaneous production of three $J/\psi$ mesons in proton proton collisions \cite{CMS:2021qsn}.
LHCb collaboration also very recently measured the tetra quark states with the quark contents of doubly charmed \cite{LHCb:2019} and doubly charmed - anticharmed states \cite{LHCb:2020bwg}.
Further details on the recent development on exotic states on experimental as well as on the theoretical front can be found in the review articles \cite{Liu:2019zoy,Brambilla:2019esw,Faustov:2021hjs}.

Within the open flavour threshold, these states provide excellent opportunity for testing different theoretical approaches including effective field theories.
Some of the recent approaches include attempts based on first principle such as lattice quantum chromodynamics \cite{Meng:2019lkt,Liu:2020qfz,Lewis:2012ir,Wurtz:2015mqa} and
QCD sum rules \cite{Aliev:2019wcm,Azizi:2017izn}.
The others include perturbative QCD in which the bottomonium spectrum is computed perturbatively at next-to-next-to-next-to-leading order \cite{Kiyo:2013aea}.
Na\"ive Cornell potential was also calibrated for lowest lying bottomonium states upto next-to-next-to-next-to-leading order using nonrelativistic QCD formalism \cite{Mateu:2018zym}.
Various decay widths and electromagnetic productions of heavy quarkonia are also computed within the heavy quark effective field theory of potential nonrelativistic QCD formalism \cite{Brambilla:2020xod}.
$B_c$ meson mass spectra has been computed using Dyson-Schwinger and Bethe-Salpeter equation approach \cite{Chen:2020ecu},  nonrelativistic as well as relativistic quark models \cite{Ortega:2020uvc,Li:2019tbn,Ebert:2011jc}.
Bottomonium mass spectrum has been obtained employing relativistic flux tube model \cite{Chen:2019uzm}.
Mass spectra and Regge trajectories have also been reported using different approaches in Refs.  \cite{Bakker:2019ynk,Chen:2018hnx}.
Spectroscopy and decay properties of heavy quarkonia and $B_c$ mesons have been studied within potential model formalism considering various confinement schemes \cite{Molina:2020zao,Eichten:2019hbb,Eichten:2019gig,Pandya:2021vby,Chaturvedi:2019usm,Chaturvedi:2020skw,Chaturvedi:2018xrg,Deng:2016ktl,Deng:2016stx,Segovia:2016xqb,Bhavsar:2018umj,Pandya:2014}.

Many of the above phenomenological studies have successfully computed mass spectra while the others have been successful in computation of decay properties. Also, the number of model parameters used vary with systems in most studies. This has motivated us to have a comprehensive study of mass spectra, decays and other related properties of heavy quarkonia and $B_c$ mesons with the least number of parameters. We complement our earlier study \cite{Soni:2017wvy} in this paper and provide the wave functions, annihilation widths of heavy quarkonia as well as scalar charge radii of $B_c$ meson without any additional parameter.
In our earlier work \cite{Soni:2017wvy},  we had made the comprehensive study of heavy quarkonia and  $B_c$ meson in the nonrelativistic potential model framework using the Cornell potential.
With the help of model parameters and numerical wave function, we have computed various decay properties such as leptonic decay constants, digamma, digluon, dilepton and electromagnetic transitions widths.
This work will provide information regarding the wave functions,  scalar charge radii,  some weak decay properties as well as the annihilation widths for heavy quarkonia.

This paper is organised in the following way. After the brief information regarding the recent theoretical progress on heavy quarkonium and $B_c$ meson, in Sec \ref{sec:formalism}, we shortly outline the nonrelativistic Cornell potential model and also provide the numerical wave functions of ground state as well as the excited states.
Then in Sec \ref{sec:decays}, we provide the formulation for weak decays using the nonrelativistic Van Royen Weiskopf formula and in Sec. \ref{sec:results}, we provide all the results for the annihilation widths. Finally we summarise the work presented here.

\section{Formalism}
\label{sec:formalism}
\begin{figure*}[htbp]
\begin{center}
\includegraphics[width=0.4\textwidth]{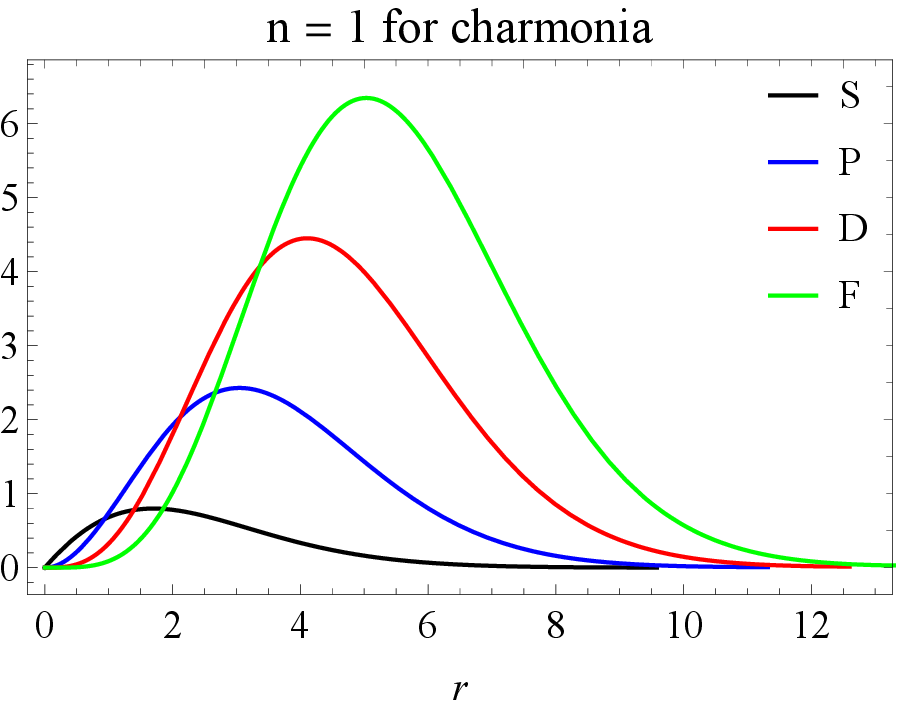}
\hfill\includegraphics[width=0.42\textwidth]{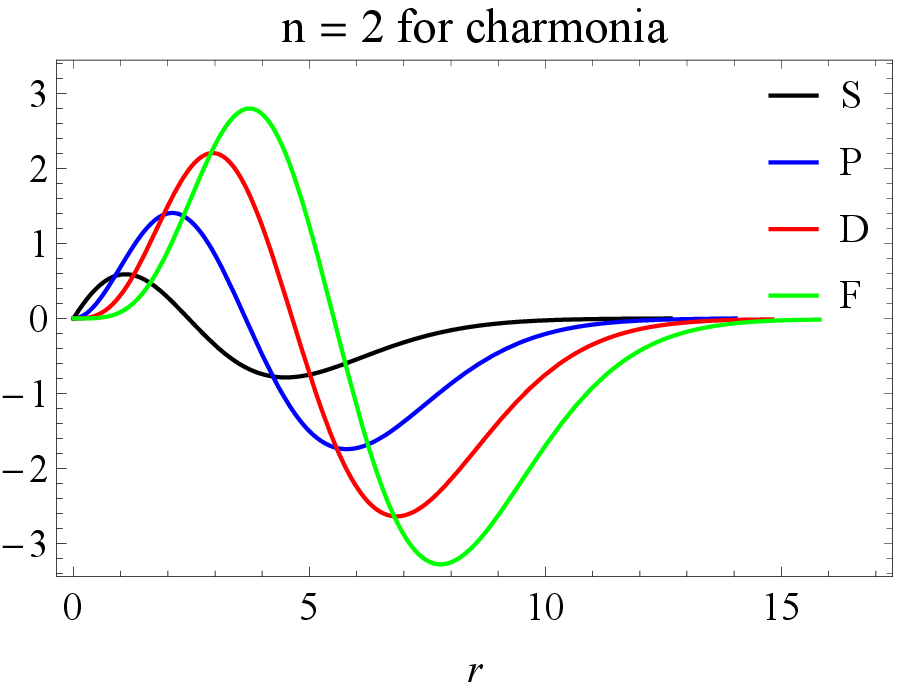}\\
\vspace{0.15in}
\includegraphics[width=0.4\textwidth]{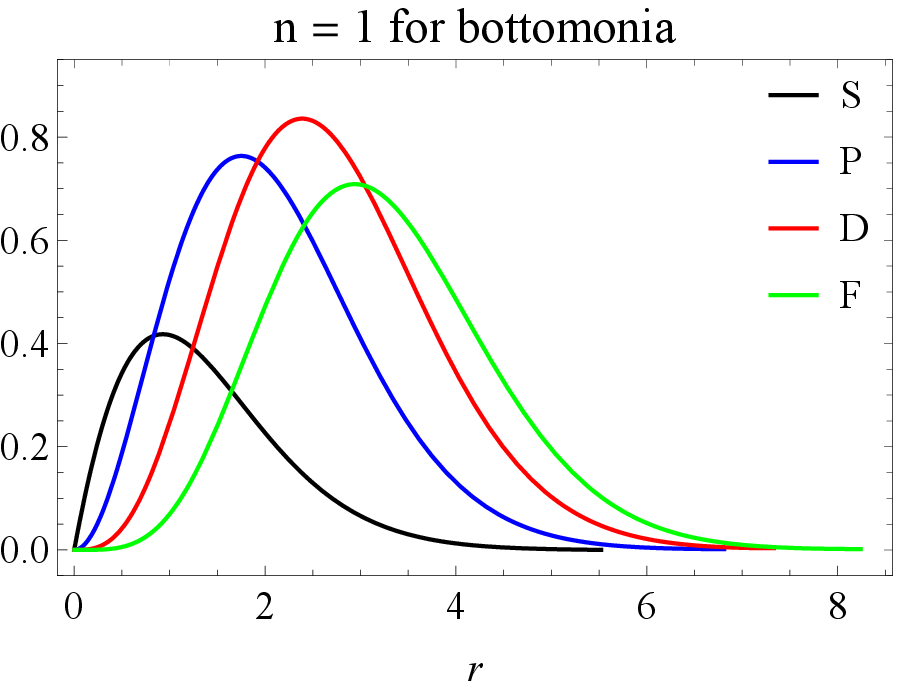}
\hfill\includegraphics[width=0.42\textwidth]{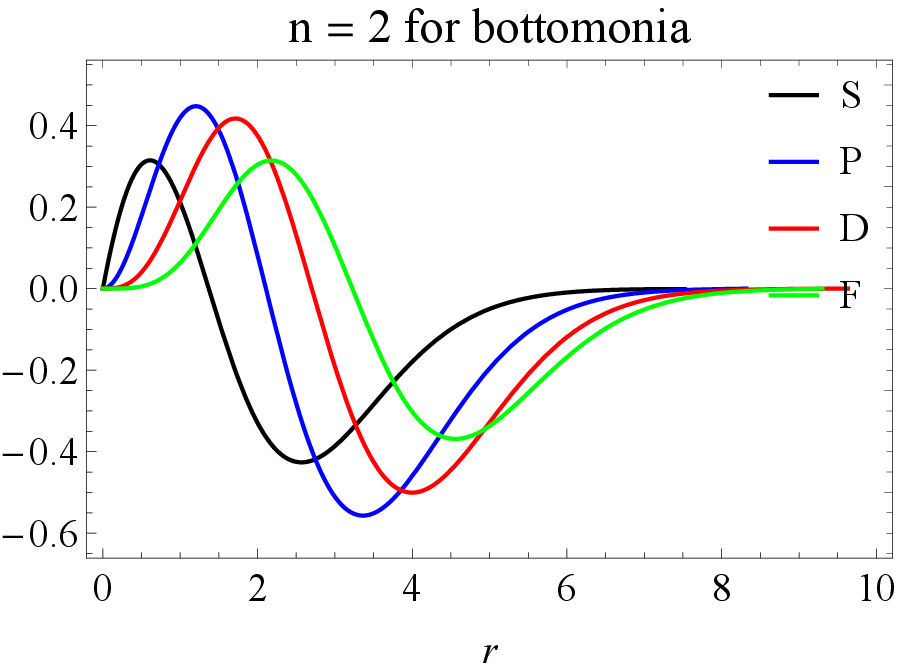}\\
\vspace{0.15in}
\includegraphics[width=0.4\textwidth]{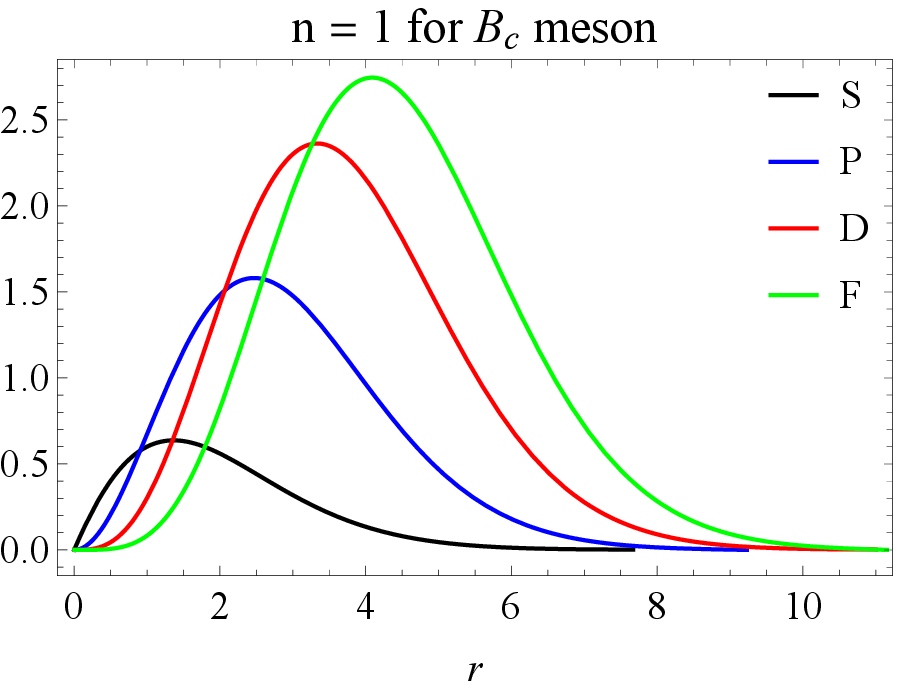}
\hfill\includegraphics[width=0.42\textwidth]{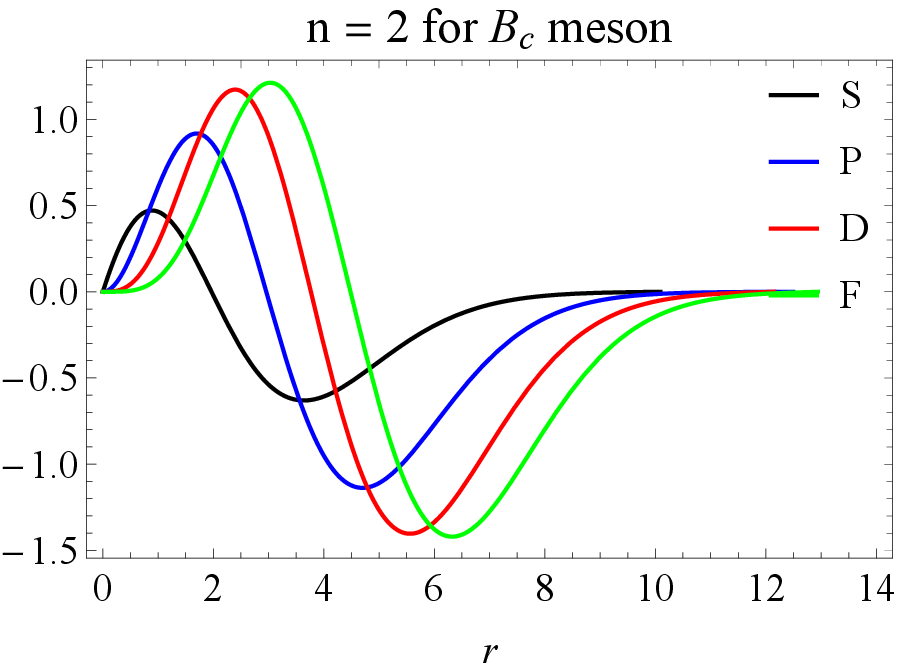}
\caption{Reduced wave function for charmonia,  bottomonia and $B_c$ mesons}
\label{fig:wavefunction}
\end{center}
\end{figure*}
For any theoretical model, it is important to predict the decay properties along with the mass spectra. Many theoretical attempts correctly predict the mass spectra but fail to reproduce the decay properties in accordance with the experiments.
For computing the mass spectra of the heavy quarkonia, we choose the linear confinement along with  Coulombic interaction, namely Cornell potential given by \cite{Eichten:1974af,Eichten:1975ag,Eichten:1978tg},
\begin{eqnarray}
\label{eq:Cornell}
V(r) = - \frac{4}{3} \frac{\alpha_s}{r} + A r
\end{eqnarray}
where $\alpha_s$ is the strong running coupling constant and $A$ is the confinement strength. The Cornell potential serves as the most prominent and widely accepted potential for nonrelativistic treatment of bound states and is also supported by lattice calculations.
In our earlier work \cite{Soni:2017wvy}, we had successfully employed the Cornell potential for computation of mass spectra of heavy quarkonia and $B_c$ meson along with a range of decay properties with least number of independent model parameters, namely quark masses and confinement strengths of corresponding quarkonia states.
For computation of mass spectra, we solve the Schr\"odinger equation numerically for Cornell potential Eq. (\ref{eq:Cornell}) using the Runge Kutta method utilized in  \textit{Mathematica} notebook \cite{Lucha:1998}. The model parameters are fixed by fitting the computed ground state masses with the corresponding experimental data. In order to compute the mass spectra of higher excited states, we add the spin dependent part of confined one gluon exchange potential perturbatively given by \cite{Voloshin:2007},
\begin{eqnarray}\label{eq:vsd}
V_{SD} (r) &=& V_{SS} (r) \left[S(S+1) - \frac{3}{2}\right] + V_{LS}(r) (\vec L\cdot\vec S) \nn &+& V_T(r) \left[S(S+1)-3(S\cdot \hat r) (S\cdot \hat r)\right],
\end{eqnarray}
where,  the coefficients of spin-spin, spin orbit and tensor interaction terms which provides hyper-fine and fine structure of heavy quarkonium states given as
\begin{eqnarray}
V_{SS} (r) &=& \frac{1}{3 m_Q m_{\bar Q}} \nabla^2 V_V(r) = \frac{16 \pi \alpha_s}{9 m_Q m_{\bar Q}} \delta^3(\vec{r}), \nn
V_{LS} (r) &=& \frac{1}{2 m_Q m_{\bar Q} r} \left(3 \frac{dV_V(r)}{dr} - \frac{dV_S(r)}{dr}\right), \\
V_T(r) &=& \frac{1}{6 m_Q m_{\bar Q}} \left(3 \frac{dV^2_V(r)}{dr^2} -\frac{1}{r} \frac{dV_V(r)}{dr}\right), \nonumber
\end{eqnarray}
with $V_V (r)$ and $V_S(r)$ are the vector and scalar part of the Cornell potential in Eq. (\ref{eq:Cornell}) respectively.
Using the potential parameters and numerical wave function, we have computed various annihilation widths including digamma, digluon, dilepton and electromagnetic ($E1$ and $M1$) transitions.
We have also computed the $B_c$ spectroscopy using the parameters used for charmonium and bottomonium spectroscopy and our results of mass spectra are in excellent agreement with the  recent data from LHCb \cite{Aaij:2020jrx}. Our results match with the ground as well as the excited state masses from LHCb \cite{Aaij:2019ldo,Aaij:2016qlz,Aaij:2014asa,Aaij:2013gia} and CMS data \cite{Sirunyan:2019osb}. The potential parameters used for the computation are $m_c = 1.317~GeV$, $m_b = 4.584~GeV$, $A_{c\bar{c}} = 0.18~GeV^2$,  $A_{b\bar{b}} = 0.25~GeV^2$ and $A_{c\bar{b}} = 0.215~GeV^2$  \cite{Soni:2017wvy}.

In this article, we provide the numerical values of the wave function of ground state as well as radial excited states at zero quark antiquark separation. This can be computed using the relation
\begin{equation}
R_{n\ell}^\ell (0) = \left. \frac{d^\ell R_{n\ell}(r)}{dr^\ell}\right |_{r~=~0}.
\end{equation}
The wave function utilised for computation of various decay widths is mentioned above.
For computation of the wave functions for different $nJ$ states, we use the relations given in the Ref.  \cite{Rai:2008prc,Rai:2008} as,
\begin{eqnarray}
R_{nJ} (0) = R(0) \left[1 + (\mathrm{SF})_J \frac{\langle\varepsilon_{SD}\rangle _{nJ}}{M_{\mathrm{SA}}}\right].
\end{eqnarray}
Here,  $R(0)$ is the radial wave function at zero quark antiquark separation,  $M_{\mathrm{SA}}$ is the spin average mass,  $(\mathrm{SF})_J$ is the spin factor and $\langle\varepsilon_{SD}\rangle _{nJ}$ is the spin interaction energy in the respective $J$ state.
Using the same potential parameters, we also study some additional annihilation widths such as digluon and dilepton decay width of $D$-wave quarkonia, $3\gamma$, $3g$ and $\gamma gg$ decay widths of heavy quarkonia without using any additional parameter.
\begin{figure*}[htbp]
\begin{center}
\includegraphics[width=0.4\textwidth]{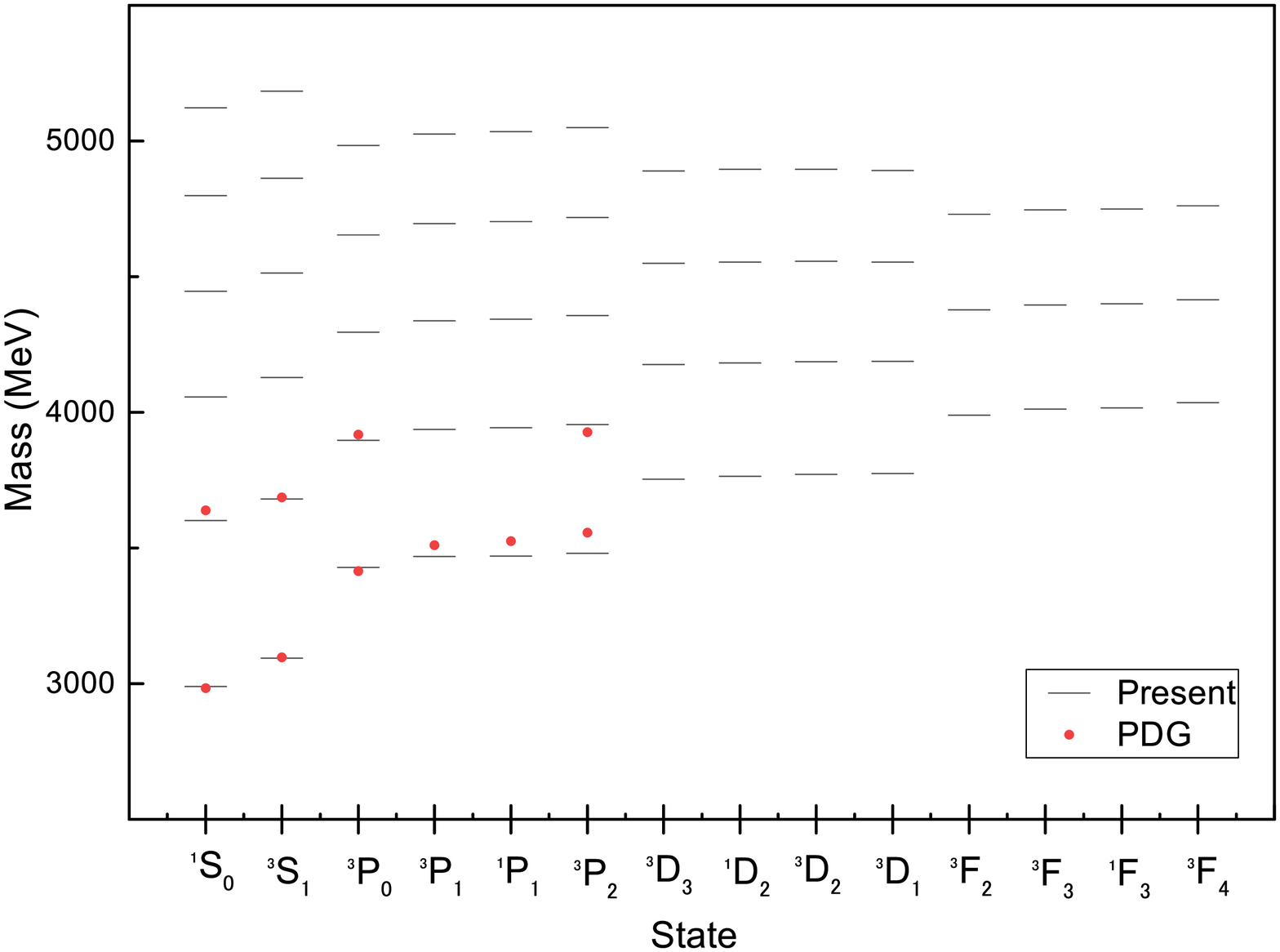}
\hfill\includegraphics[width=0.4\textwidth]{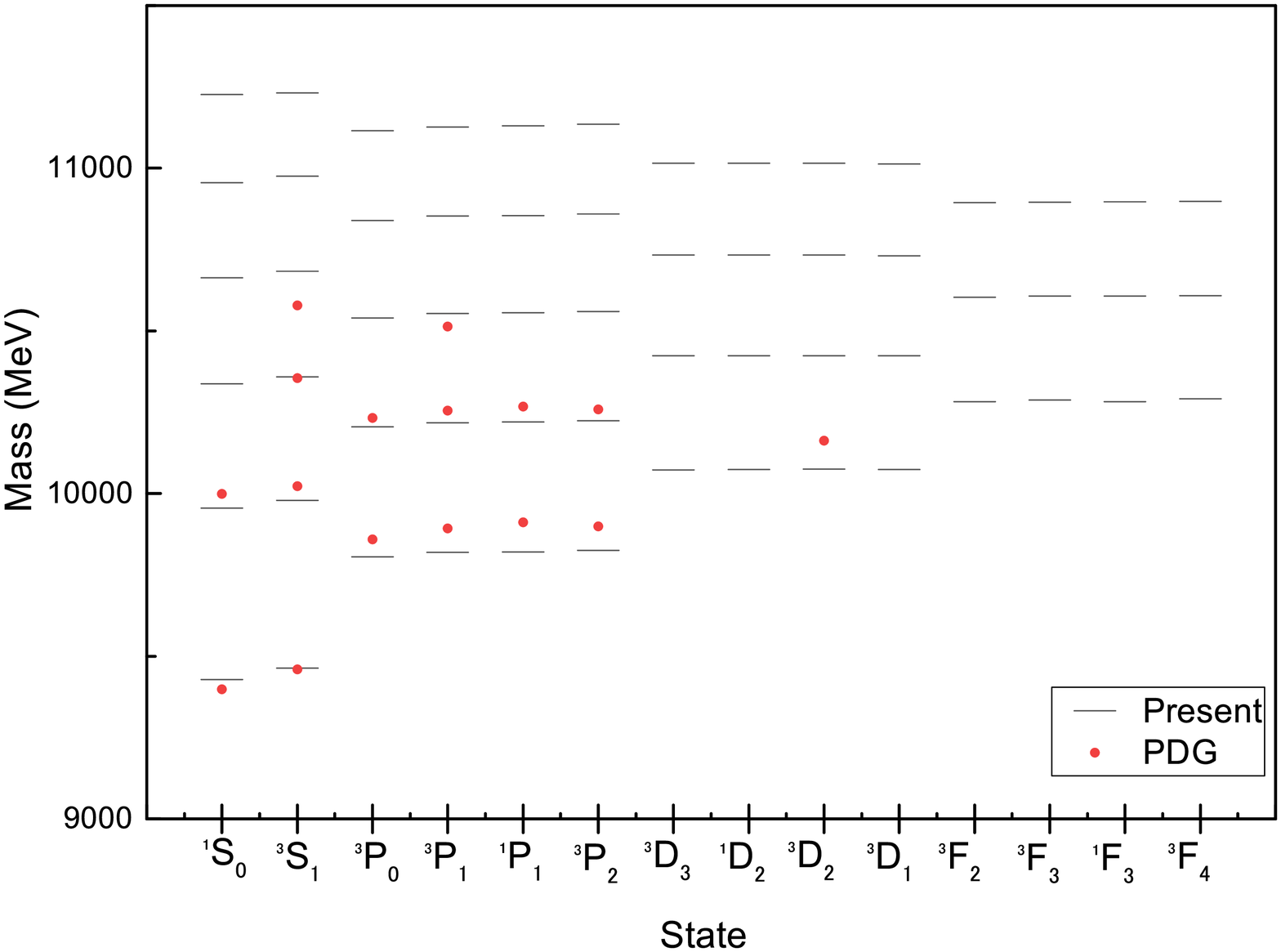}\\
\includegraphics[width=0.4\textwidth]{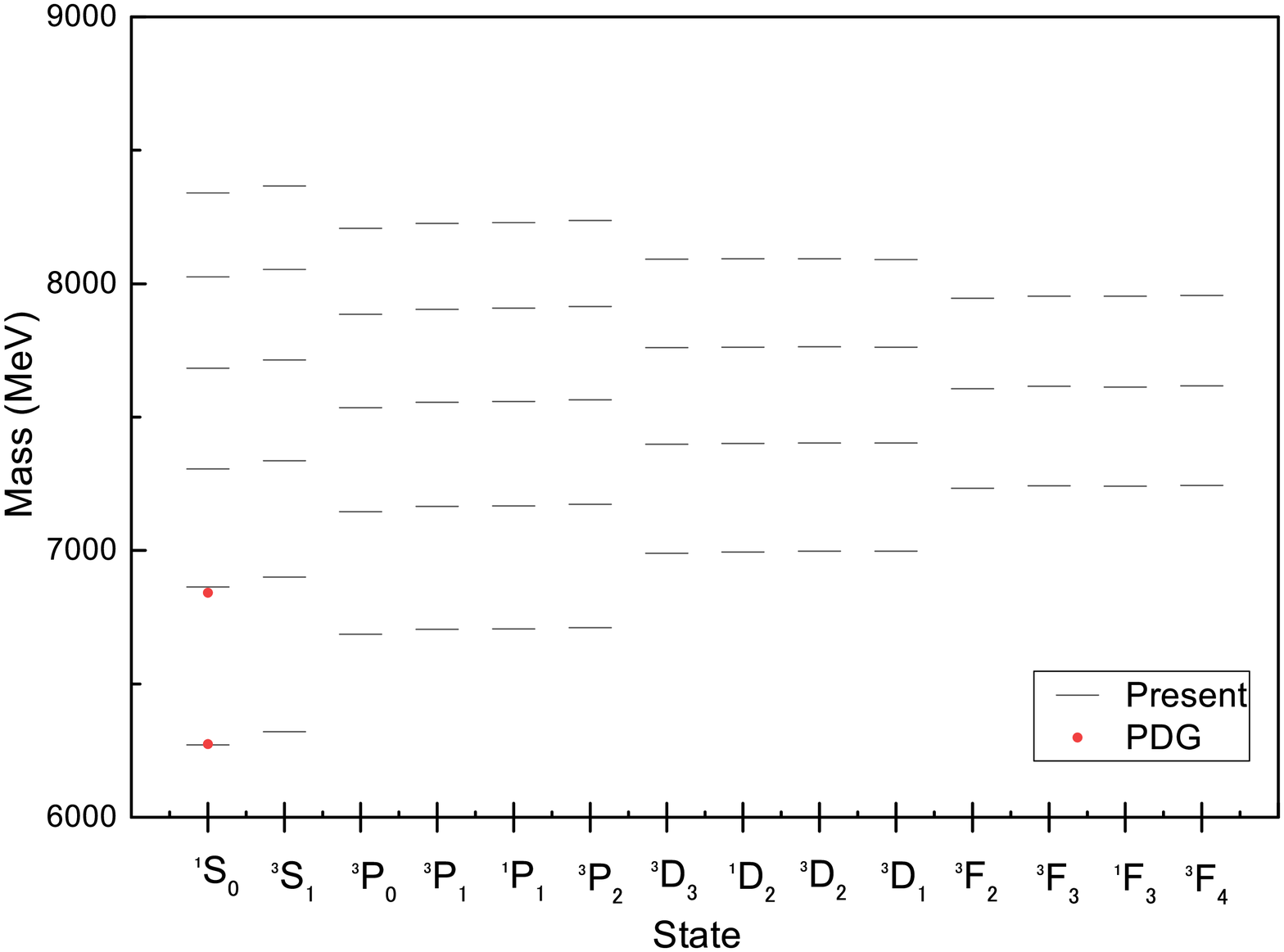}
\caption{Mass spectrum of charmonia (upper left), bottomonia (upper right) and $B_c$ mesons (bottom).}
\label{fig:qq}
\end{center}
\end{figure*}
\section{Computation of annihilation widths}
\label{sec:decays}
Annihilative decays serve as the important probe for understanding the heavy quark dynamics within quarkonia. Also, they have major contribution to the total decay widths.
Since these decays depend on the wave function at origin, it ultimately tests the validity of the potential.
Using the potential parameters and wave function, we compute various decay properties as follows.
\subsection{$Q\bar{Q}(n^3S_1 \to (3\gamma, 3g, \gamma gg))$ decay width}
In the nonrelativistic limit, the annihilation widths of heavy quarkonia are directly proportional to $|R_{n\ell}^\ell(0)|^2$. The annihilation widths of $S$-wave vector quarkonia into gluons and photons with first order QCD radiative correction are expressed as \cite{VanRoyen:1967,Kwong:1988,Kwong:1988ae}
\begin{eqnarray}
\Gamma\left(n^{3}S_{1}\rightarrow3\gamma\right)&=&\frac{4 (\pi^{2}-9)e_{Q}^{6}\alpha^{3}\mid R_{nsV}\left(0\right)\mid^{2}}{3\pi m_{Q}^{2}} \left(1-\frac{12.6\alpha_{s}}{\pi}\right)\nonumber
\\
\Gamma\left(n^{3}S_{1}\rightarrow \gamma gg\right)&=&\frac{8 (\pi^{2}-9)e_{Q}^{2}\alpha\alpha_{s}^{2}\mid R_{nsV}\left(0\right)\mid^{2}}{9\pi m_{Q}^{2}} \left(1-\frac{C_0\alpha_{s}}{\pi}\right)\nonumber
\\
\Gamma\left(n^{3}S_{1}\rightarrow3g\right)&=&\frac{10 (\pi^{2}-9)\alpha_{s}^{3}\mid R_{nsV}\left(0\right)\mid^{2}}{81\pi m_{Q}^{2}}\left(1-\frac{C_{0Q}\alpha_{s}}{\pi}\right)\nonumber
\end{eqnarray}
where the bracketed terms are the QCD first order radiative corrections. Also $R_{nsV} (0)$ is the wave function at the origin of vector state of $S$-wave quarkonia.
Also $e_Q$ is the charge of heavy quark, $\alpha_s$ is the strong running coupling constant and $\alpha$ is the electromagnetic coupling constant.
The coefficients $C_0$ and $C_{0Q}$ have values 6.7, 3.7 for charmonia and 7.4, 4.9 for bottomonia, respectively.
The annihilation widths of $S$-wave vector quarkonia to $3\gamma$, $3g$ and $\gamma gg$ are tabulated in Tab. \ref{tab:3gamma}, \ref{tab:3g}, \ref{tab:gamma_gg}.

\subsection{$Q\bar{Q}(n^1P_1 \to 3g)$ decay width}
The annihilation widths $P$-wave to gluons state given by \cite{Kwong:1988,Kwong:1988ae} ,
\begin{eqnarray}
\Gamma\left(n^{1}P_{1}\rightarrow3g\right)&=&\frac{20 \alpha_{s}^{3}\mid R_{nP}^{\prime}\left(0\right)\mid^{2}}{9\pi m_{Q}^{4}}ln(m_{Q}\langle r \rangle)\nonumber
\end{eqnarray}
Note that for $P$-wave, the decay rates are proportional to the first derivative of the wave function.

\subsection{$Q\bar{Q}(n^3D_1 \to e^+e^-)$, $Q\bar{Q}(n^1D_2 \to gg)$ and $Q\bar{Q}(n^3D_J \to 3g)$ decay width}
Similarly, the decay widths of $D$-wave heavy quarkonia to dilepton, digluon and three gluon are given by \cite{Segovia:2016xqb}. Here, the decay widths are proportional to the second derivative of the wave function.
\begin{eqnarray}
\Gamma\left(n^{3}D_{1}\rightarrow e^{+}e^{-}\right)&=&\frac{25e_{Q}^{2}\alpha^{2}\mid R_{nD}^{\prime\prime}\left(0\right)\mid^{2}}{2m_{Q}^{4}M_{nD}^{2}}\nn
\Gamma\left(n^{1}D_{2}\rightarrow gg\right)&=&\frac{2\alpha_{s}^{2}\mid R_{nD}^{\prime\prime}
\left(0\right)\mid^{2}}{3\pi m_{Q}^{6}}\nn
\Gamma\left(n^{3}D_{1}\rightarrow3g\right)&=&\frac{760 \alpha_{s}^{3}\mid R_{nP}^{\prime\prime}\left(0\right)\mid^{2}}{81\pi m_{Q}^{6}}ln(4m_{Q}\langle r \rangle)\nn
\Gamma\left(n^{3}D_{2}\rightarrow3g\right)&=&\frac{10 \alpha_{s}^{3}\mid R_{nP}^{\prime\prime}\left(0\right)\mid^{2}}{9\pi m_{Q}^{4}}ln(4m_{Q}\langle r \rangle)\nn
\Gamma\left(n^{3}D_{3}\rightarrow3g\right)&=&\frac{40 \alpha_{s}^{3}\mid R_{nP}^{\prime\prime}\left(0\right)\mid^{2}}{9\pi m_{Q}^{6}}ln(4m_{Q}\langle r \rangle)\nonumber
\end{eqnarray}
The annihilation widths are listed in Tab. \ref{tab:3gamma} - \ref{tab:digluon}.  
Note here that in some of annihilation rates, the correction factors are not available for the higher order excited states. We compare our findings with the available PDG data as well as theoretical predictions.

\section{Results and Discussion}
\label{sec:results}
In this article, we have computed dilepton, digluon, $3\gamma$, $3g$ and $\gamma gg$ annihilation widths of respective heavy quarkonia. For computations of these decay widths, we have employed the respective masses and input parameters from our earlier paper \cite{Soni:2017wvy}.
For computation of mass spectra, we solve the Schr\"odinger equation numerically for the Cornell potential Eq. (\ref{eq:Cornell}) and the spin dependent parts of confined one gluon exchange potential are treated perturbatively for the ground as well as excited states.
The computed masses for charmonia, bottomonia and $B_c$ mesons are shown in Fig. \ref{fig:qq} in comparison with experimental data.
In Tab. \ref{tab:wavefunction}, we provide the absolute square of radial wave function at zero quark - antiquark separation which utilised for computations of the annihilation decay widths.
The computed wave functions can also be utilised for computing the heavy quarkonium production cross sections.
Using the numerical wave function, we also compute the scalar charge radii for the ground state as well as for the excited states and tabulate them in Tab. \ref{tab:wavefunction}, which can be used for computations of hadronic transition widths.
Note that these charge radii gives the information regarding the charge distribution within the hadron and thus helpful in computation of the electromagnetic form factors.
For computation of scalar charge radii, we use the general method given in Ref. \cite{Vinodkumar:1999,Pandya:2001}.
It is expected that the charge radii increases for higher excited states as the contribution of confinement term dominates. The same can be observed in the data presented in Tab.  \ref{tab:wavefunction}.
Using the potential parameters, the weak annihilation widths are computed using the leading order nonrelativistic relations including the available QCD corrections.
We compare our results with available experimental data as well as different theoretical approaches. These theoretical approaches include nonrelativistic constituent quark model (NRCQM) for bottomonium spectrum \cite{Segovia:2016xqb}, potential nonrelativistic quantum chromodynamics (pNRQCD) formalism in which authors have computed mass spectra and decay properties employing relativistic corrections to the Cornell potential \cite{Chaturvedi:2019usm}.
We also compare our results with different potential model estimations which include the Schr\"odinger formalism using Cornell potential for charmonia \cite{Chaturvedi:2018xrg,Kher:2018wtv}, relativistic Dirac formalism with linear confinement \cite{Bhavsar:2018umj} as well as instanton induced quarkonia potential obtained from instanton liquid model for QCD vacuum \cite{Pandya:2021vby}.

We present the non-normalised reduced wave functions of ground state as well as excited states for charmonia, bottomonia and $B_c$ mesons in Fig. \ref{fig:wavefunction}.
In Tab. \ref{tab:3gamma}, we provide the results of $3\gamma$ decay widths of charmonia and bottomonia along with different potential model estimations and nonrelativistic constituent quark model results. For charmonia, our results are matching well with different potential model estimations and pNRQCD potential model. Our results are also very near to the world average of experimental data from CLEO \cite{Besson:2008pr} and BESIII data \cite{Ablikim:2012zwa}. For bottomonia, the $3\gamma$ decay width is highly suppressed and still the experimental results are not available. Our results are found to be between those reported using NRCQM and other potential model results.

In Tab. \ref{tab:3g}, we provide the $3g$ decay widths from various quarkonium states. For $\Gamma(n^3S_1 \to 3g)$, our decay widths overestimate the experimental data. For Charmonia, our results are consistent with the other potential model estimation for the channel $\Gamma(J/\psi \to 3g)$ and for bottomonia our results are matching quite well with other potential model estimation and NRCQM results. For $P$ and $D$-wave charmonia, our results are an order higher than other potential model estimation but for bottomonia, our results are consistent with the other potential models estimation and NRCQM results.
The discrepancy in comparison with the other potential model estimation \cite{Kher:2018wtv} is due to different formalisms utilised.  In Ref.  \cite{Kher:2018wtv}, the authors have used the Gaussian trial wave function for solving the Schr\"odinger equation for the Cornell potential whereas in our computation, we solve Schr\"odinger equation numerically.

In Tab. \ref{tab:gamma_gg}, we provide the decay widths of $n^3S_1 \to \gamma gg$ and it is observed that our results are matching fairly well with experimental data,  other potential model estimations as well as NRCQM results.
In Tab.  \ref{tab:dilepton},  we provide the dilepton decay widths of charmonia and bottomonia.  Note that for this computation,  the first order QCD correction is also given in Ref.  \cite{Bradley:1980eh},  however,  we do not consider here in the present work.  Further,  in the NRCQM studies also,  this first order correction is not included in decay width computation of this channel \cite{Segovia:2016xqb}.
For charmonia,  our results are near to the other potential model estimation. However, our results are consistently lower than the other references for bottomonia.
Lastly,  in Tab.  \ref{tab:digluon}, we provide the digluon decay widths of $D$-wave and charmonia and bottomonia along with the results from different other potential model estimations.
It is interesting to note that for charmonia, our results are consistently higher than the other potential model  estimation, however for bottomonia,  our results are in good agreement with the NRCQM studies.
It is to be noted here that for the annihilation widths of $P$ and $D$ states to gluons, we have not considered any radiative corrections.  Further,  these correction factors are significant particularly for open flavour mesons.  The contribution of these correction factors are also very small for charmonia and bottomonia  being heavy states.

\begin{table*}
\centering
\caption{\textbf{Absolute square of spin averaged radial wave function at origin ($|R_{nl}^{(l)} (0)|^2$ in unit of GeV$^{2\ell+3}$) and scalar charge radii ($\sqrt{\langle r^2 \rangle}$ in unit of fm) for charmonia, bottomonia and $B_c$ mesons.}}
{\begin{tabular}{@{}c|cc|cc|cc@{}}
\hline\hline
State & \multicolumn{2}{c|}{$c\bar{c}$} & \multicolumn{2}{c|}{$c\bar{b}$} & \multicolumn{2}{c}{$b\bar{b}$}\\
\cline{2-7}
& $|R_{nl}^{(l)} (0)|^2$ & $\sqrt{\langle r^2 \rangle}$& $|R_{nl}^{(l)} (0)|^2$ & $\sqrt{\langle r^2 \rangle}$& $|R_{nl}^{(l)} (0)|^2$ & $\sqrt{\langle r^2 \rangle}$\\
\hline
$1S$ & 0.6396  	& 0.463 & 1.2457	& 0.373 & 4.1438 	& 0.259\\
$2S$ & 0.4775  	& 0.887 & 0.9156	& 0.719 & 2.8162 	& 0.512\\
$3S$ & 0.4282 	& 1.231 & 0.8167 	& 0.999 & 2.4496 	& 0.716\\
$4S$ & 0.4024  	& 1.532 & 0.7653	& 1.245 & 2.2641 	& 0.895\\
$5S$ & 0.3859  	& 1.806 & 0.7325	& 1.468 & 2.1474 	& 1.058\\
$6S$ & 0.3741 	& 2.061 & 0.7091	& 1.676 & 2.0652 	& 1.209\\
$1P$ & 0.0934  	& 0.706 & 0.1753	& 0.572 & 0.4867 	& 0.407\\
$2P$ & 0.0948 	& 1.080 & 0.1773	& 0.877 & 0.4859 	& 0.629\\
$3P$ & 0.0947  	& 1.398 & 0.1770	& 1.136 & 0.4817 	& 0.818\\
$4P$ & 0.0958 	& 1.682 & 0.1763	& 1.368 & 0.4778 	& 0.986\\
$5P$ & 0.0944  	& 1.945 & 0.1757	& 1.581 & 0.4747 	& 1.141\\
$1D$ & 0.0291  & 0.902 & 0.8626	& 0.733 & 0.4382 	& 0.525\\
$2D$ & 0.0486 	& 1.246 & 0.1376	& 1.013 & 0.7248 	& 0.729\\
$3D$ & 0.0646  & 1.546 & 0.1827	& 1.257 & 0.9577 	& 0.907\\
$4D$ & 0.0786  & 1.819 & 0.2222	& 1.479 & 1.1610 	& 1.068\\
$1F$ & 1.1571 $\times 10^{-5}$		& 1.077 & 7.5916 $\times 10^{-5}$	& 0.875 & 0.0016 	& 0.630\\
$2F$ & 3.4380 $\times 10^{-5}$		& 1.399 & 2.2501 $\times 10^{-4}$ 	& 1.137 & 0.0046 	& 0.820\\
$3F$ & 6.7986 $\times 10^{-5}$ 		& 1.685 & 4.4413 $\times 10^{-4}$	& 1.370 & 0.0090 	& 0.989\\
$4F$	 & 11.2058 $\times 10^{-5}$	& 1.949 & 7.3102 $\times 10^{-4}$	& 1.585 & 0.0147 	& 1.145\\
\hline\hline
\end{tabular}\label{tab:wavefunction}}
\end{table*}
\begin{table*}
\centering
\caption{$3\gamma$ decay widths of charmonia (in eV) and bottomonia (in $10^{-6}$ keV)}
{\begin{tabular}{@{}lccccc|lccc@{}}
\hline\hline
& & pNRQCD & PM & PM & 	PDG && &PM& NRCQM\\
State & Present & \cite{Chaturvedi:2019usm} & \cite{Kher:2018wtv} & \cite{Chaturvedi:2018xrg} &\cite{PDG:2018}& State & Present & \cite{Pandya:2021vby} & \cite{Segovia:2016xqb} \\
\hline
$J/\psi$		& 1.36	& 1.02 & 3.95	& 2.997 & $1.08 \pm 0.032$ 	& $\Upsilon(1S)$ & 7.05 & 123.12 & 3.44\\
$\psi(2S)$	& 1.01	& 0.90 & 1.64	& 1.083 &	--							& $\Upsilon(2S)$ & 4.79 & 89.89 & 2.00\\
$\psi(3S)$	& 0.91	& 0.86 & 1.39	& 1.046 & --							& $\Upsilon(3S)$ & 4.16 & 72.04 & 1.55\\
$\psi(4S)$	& 0.85	& 0.83 & 1.30	& 0.487 & --							& $\Upsilon(4S)$ & 3.85 & 60.34 & 1.29\\
$\psi(5S)$	& 0.81	& 0.81 & 1.25	& 0.381 & --							& $\Upsilon(5S)$ & 3.64 & 52.02 & --\\
$\psi(6S)$	& 0.79	& 0.80 & 1.22	& 0.312 & --							& $\Upsilon(6S)$ & 3.51 & 44.93 & --\\
\hline\hline
\end{tabular}\label{tab:3gamma}}
\end{table*}
\begin{table*}
\centering
\caption{$3g$ decay widths of charmonia (in keV) and bottomonia (in keV)}
{\begin{tabular}{@{}lccc|lcccc@{}}
\hline\hline
& & PM & 	PDG & & & PM & NRCQM & PDG\\
State & Present & \cite{Kher:2018wtv} & \cite{PDG:2018}& State & Present & \cite{Pandya:2021vby} & \cite{Segovia:2016xqb} & \cite{PDG:2018}\\
\hline
$J/\psi$		& 264.25	& 269.06	& 59.55	& $\Upsilon(1S)$ 	& 39.15	& 40.0 & 41.63 	& --\\
$\psi(2S)$	& 196.05	& 112.03	& 31.16	& $\Upsilon(2S)$ 	& 26.59	& 26.9 & 24.25 	& 18.80\\
$\psi(3S)$	& 175.43	& 94.57		& --		& $\Upsilon(3S)$ 	& 23.13	& 20.6 & 18.76 	& 7.25\\
$\psi(4S)$	& 164.66	& 88.44		& 	--		& $\Upsilon(4S)$ 	& 21.37	& 16.8 & 15.58 	& --\\
$\psi(5S)$	& 157.77	& 85.30		& --		& $\Upsilon(5S)$ 	& 20.27	& 14.1 & --			& --\\
$\psi(6S)$	& 152.86	& 83.19		& --		& $\Upsilon(6S)$ 	& 19.49	& 11.7 & --			& --\\
$h_c(1P)$	& 1129.11	& 285.127	&			& $h_b(1P)$			& 14.91	& 35.7 & 35.26		& \\
$h_c(2P)$	& 1459.74	& 420.078	&			& $h_b(2P)$			& 17.76	& 34.6 & 52.70		& \\
$h_c(3P)$	& 1649.05	& 558.780	&			& $h_b(3P)$			& 19.33	& 33.1 & 62.14		& \\
$h_c(4P)$	& 2098.68	&	--			&			& $h_b(4P)$			& 20.40	& 32.7 & --			& \\
$\psi (1D)$	& 1758.46	& 189.367	& & $\Upsilon (1D)$		& 4.667	& 10.6 & 9.97\\
$\psi_2 (1D)$& 208.239	& 53.876	& & $\Upsilon_2 (1D)$	& 0.552	& -- & 0.62\\
$\psi_3 (1D)$& 832.956	& 89.700	& & $\Upsilon_3 (1D)$	& 2.211	& 6.0 & 0.22\\
$\psi (2D)$	& 3231.11	& 359.346	& & $\Upsilon (2D)$		& 8.370	& 11.9 & 9.69\\
$\psi_2 (2D)$& 382.632	& 102.236	& & $\Upsilon_2 (2D)$	& 0.991	& -- & 0.61\\
$\psi_3 (2D)$& 1530.53	& 170.217	& & $\Upsilon_3 (2D)$	& 3.965	& 5.6 & 1.25\\
$\psi (3D)$	& 4588.97	& 556.588	& & $\Upsilon (3D)$		& 11.630& 11.8 & --\\
$\psi_2 (3D)$& 539.878	& 158.353	& & $\Upsilon_2 (3D)$	& 1.377	& -- & --\\
$\psi_3 (3D)$& 2159.51	& 263.647	& & $\Upsilon_3 (3D)$	& 5.509	& 5.1 & --\\
\hline\hline
\end{tabular}\label{tab:3g}}
\end{table*}
\begin{table*}
\centering
\caption{$\gamma gg$ decay widths of charmonia (in keV) and bottomonia (in keV)}
{\begin{tabular}{@{}lccc|lcccc@{}}
\hline\hline
& & PM & 	PDG & & & PM & NRCQM & PDG\\
State & Present & \cite{Kher:2018wtv} &\cite{PDG:2018}& State & Present & \cite{Pandya:2021vby} & \cite{Segovia:2016xqb} & \cite{PDG:2018} \\
\hline
$J/\psi$		& 9.72 & 8.90	& 8.17 	& $\Upsilon(1S)$ & 0.85 & 0.72 & 0.79	& 1.18\\
$\psi(2S)$	& 7.21 & 3.75	& 3.03	& $\Upsilon(2S)$ & 0.58 & 0.49 & 0.46	& 0.60\\
$\psi(3S)$	& 6.46 & 3.16	& --		& $\Upsilon(3S)$ & 0.50 & 0.39 & 0.36	& 0.01\\
$\psi(4S)$	& 6.06 & 2.96	& --		& $\Upsilon(4S)$ & 0.46 & 0.32 & 0.30	& --\\
$\psi(5S)$	& 5.81 & 2.85	& --		& $\Upsilon(5S)$ & 0.44 & 0.27 & 0.25	& --\\
$\psi(6S)$	& 5.63 & 2.78	& --		& $\Upsilon(6S)$ & 0.42 & 0.23 & 0.22	& --\\
\hline\hline
\end{tabular}\label{tab:gamma_gg}}
\end{table*}
\begin{table*}
\centering
\caption{dilepton decay widths of charmonia (in keV) and bottomonia (in $10^{-3}$ keV)}
{\begin{tabular}{@{}lccc|lcccc@{}}
\hline\hline
State & Present & PM \cite{Bhavsar:2018umj} & PM \cite{Kher:2018wtv} & State & Present & PM \cite{Bhavsar:2018umj} & PM \cite{Pandya:2021vby} & NRCQM \cite{Segovia:2016xqb}\\
\hline
$\psi(1D)$		& 0.201  	& 0.27 		& 0.113 	& $\Upsilon (1D)$ & 0.72 & 106	& 5.0 & 1.40  \\
$\psi(2D)$ 		& 0.272	& 0.17 		& 0.166 	& $\Upsilon (1D)$ & 1.12 & 78 	& 5.8 & 2.50\\
$\psi(3D)$ 		& 0.306	& 0.099		& 0.211 	& $\Upsilon (1D)$ & 1.39 & 51 	& 5.9 & --\\
$\psi(4D)$ 		& 0.323 	& 0.064 		& -- 			& $\Upsilon (1D)$ & 1.60 & 42 	& 5.8 & --\\
\hline\hline
\end{tabular}\label{tab:dilepton}}
\end{table*}
\begin{table*}
\centering
\caption{digluon decay widths of $D$-wave charmonia (in keV) and bottomonia (in keV)}
{\begin{tabular}{@{}lcc|lcc@{}}
\hline\hline
State & Present & PM \cite{Kher:2018wtv} & State  & Present & NRCQM \cite{Segovia:2016xqb}\\
\hline
$\eta_{c2}(1D)$ & 122.194 & 12.460 	& $\eta_{b2}(1D)$ & 0.42 & 0.37 \\
$\eta_{c2}(2D)$ & 203.845 & 21.679 	& $\eta_{b2}(1D)$ & 0.69 & 0.67 \\
$\eta_{c2}(3D)$ & 270.943 & 31.757 	& $\eta_{b2}(1D)$ & 0.91 & -- \\
$\eta_{c2}(4D)$ & 329.790 & -- 			& $\eta_{b2}(1D)$ & 1.11 & -- \\
\hline\hline
\end{tabular}\label{tab:digluon}}
\end{table*}

\section{Conclusion}
In this article, we have extended our nonrelativistic treatment of heavy quarkonia for computation of annihilation to $3\gamma$, $3g$, $\gamma gg$ and $e^+e^-$. We utilise the model parameters and numerical wave function to compute these decay widths with QCD correction factors without using any additional parameters.
Our results are close to the experimental data except for the channel $3g$ decay width of charmonia.
The calculated decay widths of bottomonia are mostly consistent with the NRCQM data.
Because of the unavailability of precise experimental data and first principle studies for these channels, it is not appropriate to comment on the validity and credibility of any particular  theoretical model.
We also provide the absolute square of radial wave function at zero quark-antiquark separation and scalar charge radii for ground state as well as excited states of heavy quarkonia and $B_c$ mesons.

\section*{Acknowledgements}
J.N.P. acknowledges financial support from University Grants Commission of India under Major Research Project F.No. 42-775/2013(SR) and DST-FIST (SR/FST/PS-II/2017/20).

\section*{Conflict of interest}
Not applicable.

\clearpage
\bibliography{apssamp}
\end{document}